\documentclass[aps,prd,preprint,amsmath,amssymb]{revtex4}
\usepackage{graphicx}
\usepackage{dcolumn}
\usepackage{bm}

\newcommand{\be}{\begin{equation*}}
\newcommand{\ee}{\end{equation*}}
\newcommand{\bea}{\begin{eqnarray}}
\newcommand{\eea}{\end{eqnarray}}
\newcommand{\bean}{\begin{eqnarray*}}
\newcommand{\eean}{\end{eqnarray*}}
\newcommand{\ga}{{a^0_0(980)}}
\newcommand{\gf}{{f_0(980)}}
\newcommand{\gaf}{{$a^0_0(980)$-$f_0(980)$ mixing\ }}

\begin{document}

\title{Possibility of measuring \gaf from $J/\psi\to\phi\ga$}

\author{
Jia-Jun Wu$^{1,2}$, Qiang Zhao$^1$ and B.~S.~Zou$^{1,2}$\\
$^1$ Institute of High Energy Physics, CAS, P.O.Box 918(4), Beijing 100049, China\\
$^2$ University of Science and Technology of China, Hefei
230026,China }

\date{April 27, 2007}

\begin{abstract}
The \gaf intensity has been predicted to be in the range of 0.01to
0.1 by various theoretical models, but lacking firm experimental
observation. We examine the possibility of extracting the \gaf
from $J/\psi\to\phi\gf\to\phi\ga$ reaction at upgraded Beijing
Electron Positron Collider with BESIII detector. While the
branching ratio of this process through the \gaf is expected to be
about $O(10^{-6})$ similar to the estimated total amount from two
background reactions $J/\psi\to\gamma^{*}\to\phi\ga$ and
$J/\psi\to K^{*}\bar{K} + c.c.\to\phi\ga$, the peak width from the
\gaf is about 8 MeV, much smaller than that from other mechanisms.
With $10^9$ $J/\psi$ events at BESIII, the \gaf intensity is
expected to be unambiguously and precisely measured.
\end{abstract}


\maketitle

\section{Introduction}
\label{s1}

More than thirty years after their discovery, today the nature of
light scalar mesons $\gf$ and $\ga$ is still in controversy. They
have been described as quark-antiquark, four quarks,
$K\bar{K}$molecule, quark-antiquark-gluon hybrid, and so on. Now
the study of their nature has become a central problem in the
light hadron spectroscopy.

In the late 1970s, the mixing between the $\ga$ and $\gf$
resonances was first suggested theoretically in Ref.\cite{first}.
Its mixing intensity is expected to shed important light on the
nature of these two resonances, and has hence been studied
extensively on its different aspects and possible manifestations
in various reactions
\cite{n1,n2,n3,n5,im,uim,n4,y2,y4,y5,y10,y11,eb}. But
unfortunately no firm experimental determination on this quantity
is available yet. Only Ref.\cite{im} gives a value of
$|\xi|^{2}=(8\pm3)\%$ based on the data of the $\ga$ central
production in the reaction $pp\to p_s(\eta\pi^0)p_f$, under the
assumption that the $\ga$ resonances are mainly produced from
\gaf. However, the experimental justification of such assumption
requires measuring the reaction $pp\to p_s(\eta\pi^0)p_f$ at a
much higher energy to exclude a possible effect of the secondary
Regge trajectories, for which the $\eta\pi^0$ production is not
forbidden by G parity \cite{n1,n3,im}. Obviously, more solid and
precise measurements on this quantity are needed, such as by
polarized target experiment on the reaction $\pi^-p\to\eta\pi^0 n$
\cite{n3}, $J/\psi$ decays \cite{im}, and $dd\to\alpha\eta\pi^0$
reactions from WASA at COSY \cite{y10}.

In this paper we examine the possibility of extracting the \gaf
from $J/\psi\to\phi\ga\to\phi\eta\pi^0$ reaction. This reaction is
an isospin breaking process with initial state of isospin 0 and
final state of isospin 1. It can occur through the isospin
breaking \gaf by $J/\psi\to\phi\gf\to\phi\ga$. The
$J/\psi\to\phi\gf$ has already been clearly observed in there
action $J/\psi\to\phi\pi^+\pi^-$ by BESII experiment \cite{bes1}.
Due to poor performance for measuring multi-photon final states,
no information is available from BESII for
the$J/\psi\to\phi\eta\pi^0$ reaction, which needs to measure 4
photons from $\eta$ and $\pi^0$ decays. With $10^9$ $J/\psi$
events expected in near future at the upgraded Beijing Electron
Positron Collider (BEPCII) with much improved BESIII detector, the
measurement of the $J/\psi\to\phi\eta\pi^0$ reaction is definitely
possible. However, besides the \gaf mechanism, the
$J/\psi\to\gamma^{*}\to\phi\ga$ and $J/\psi\to K^{*}\bar{K} +
c.c.\to\phi\ga$ can also contribute to the $J/\psi\to\phi\ga$
final state. So we need to estimate relative strength of these
mechanisms to see whether we can get reliable extraction of the
\gaf.

In the next section, we give a brief review of the theory for the
\gaf and model-dependent estimations for the mixing intensity.
Then in Sect.\ref{s3} we estimate contributions of various
mechanisms to the $J/\psi\to\phi\ga$ reaction. Finally we give a
summary in Sect.\ref{s4}.

\section{Theory and estimation of $a^0_0(980)$-$f_0(980)$ mixing}
\label{s2}

The basic theory for the \gaf was already pointed out by Achasov
and collaborators \cite{first}. For the nearly degenerate $\ga$
(isospin 1) and $\gf$ (isospin 0), both can decay into $K\bar K$.
Due to isospin breaking effect, the charged and neutral kaon
thresholds are different by about 8 MeV. Between the charged and
neutral kaon thresholds the leading term to the \gaf amplitude is
dominated by the unitary cuts of the intermediate two-kaon system
and proportional to the difference of phase spaces for the charged
and neutral kaon systems.

Considering the \gaf, the propagator of $\ga/\gf$ can be expressed
as \cite{n1} :
\begin{eqnarray}
G=\frac{1}{D_{f}D_{a}-|D_{af}|^{2}}\begin{pmatrix}D_{a}&D_{af}\\D_{af}&D_{f}\end{pmatrix},
\end{eqnarray}
where $D_{a}$ and $D_{f}$ are the denominators for the usual
propagators of $\ga$ and $\gf$, respectively :
\begin{eqnarray}
\label{1}D_{a}&=&m_{a}^{2}-s- i\sqrt{s}[\Gamma^a_{\eta\pi}(s) + \Gamma^a_{K\bar K}(s)],\\
D_{f}&=&m_{f}^{2}-s- i\sqrt{s}[\Gamma^f_{\pi\pi}(s)+\Gamma^f_{K\bar K}(s)],\\
\Gamma^a_{bc}(s)&=&\frac{g^{2}_{abc}}{16\pi\sqrt{s}}\rho_{bc}(s),\\\label{2}
\rho_{bc}(s)&=&\sqrt{[1-(m_{b}-m_{c})^{2}/s][1-(m_{b}+m_{c})^{2}/s]}.
\end{eqnarray}
The  $D_{af}$ is the mixing term. From \cite{first,n3}, we have:
\begin{eqnarray}
D_{af,K\overline{K}}&=&\frac{g_{\ga K^{+}K^{-}}g_{\gf K^{+}K^{-}}}{16\pi}
\Big\{i[\rho_{K^{+}K^{-}}(s)-\rho_{K^{0}\bar{K}^0}(s)]\nonumber\\
&&-\mathcal{O}(\rho^{2}_{K^{+}K^-}(s)-\rho^{2}_{K^{0}\bar{K}^0}(s))\Big\}.\label{4}
\end{eqnarray}
The relation of $D_{af}$ and the $f_0\to a_0$ mixing parameter
$\xi$ is \cite{y4}:
\begin{eqnarray}\label{5}
|\xi|=\left|\frac{D_{af}}{D_{a}}\right|=\left|\frac{g_{\ga K^{+}K^
{-}}g_{\gf
K^{+}K^{-}}[\rho_{K^{+}K^-}(s)-\rho_{K^{0}\bar{K}^0}(s)]}{16\pi
D_{a}}\right| .
\end{eqnarray}
With the isospin breaking effect, the $\ga$ and $\gf$ meson wave
function can be expressed as  \cite{n1}:
\begin{eqnarray}
|f_{0}\rangle=cos\theta|I=0\rangle+sin\theta|I=1\rangle,\\
|a_{0}^{0}\rangle=cos\theta|I=1\rangle-sin\theta|I=0\rangle
\end{eqnarray}
with the mixing angle $\theta$ related to the mixing intensity as
$|\xi|^2 \approx sin^2\theta$.

\begin{figure}[htbp] \vspace{-0.cm}
\begin{center}
\includegraphics[width=0.37\columnwidth,angle=270]{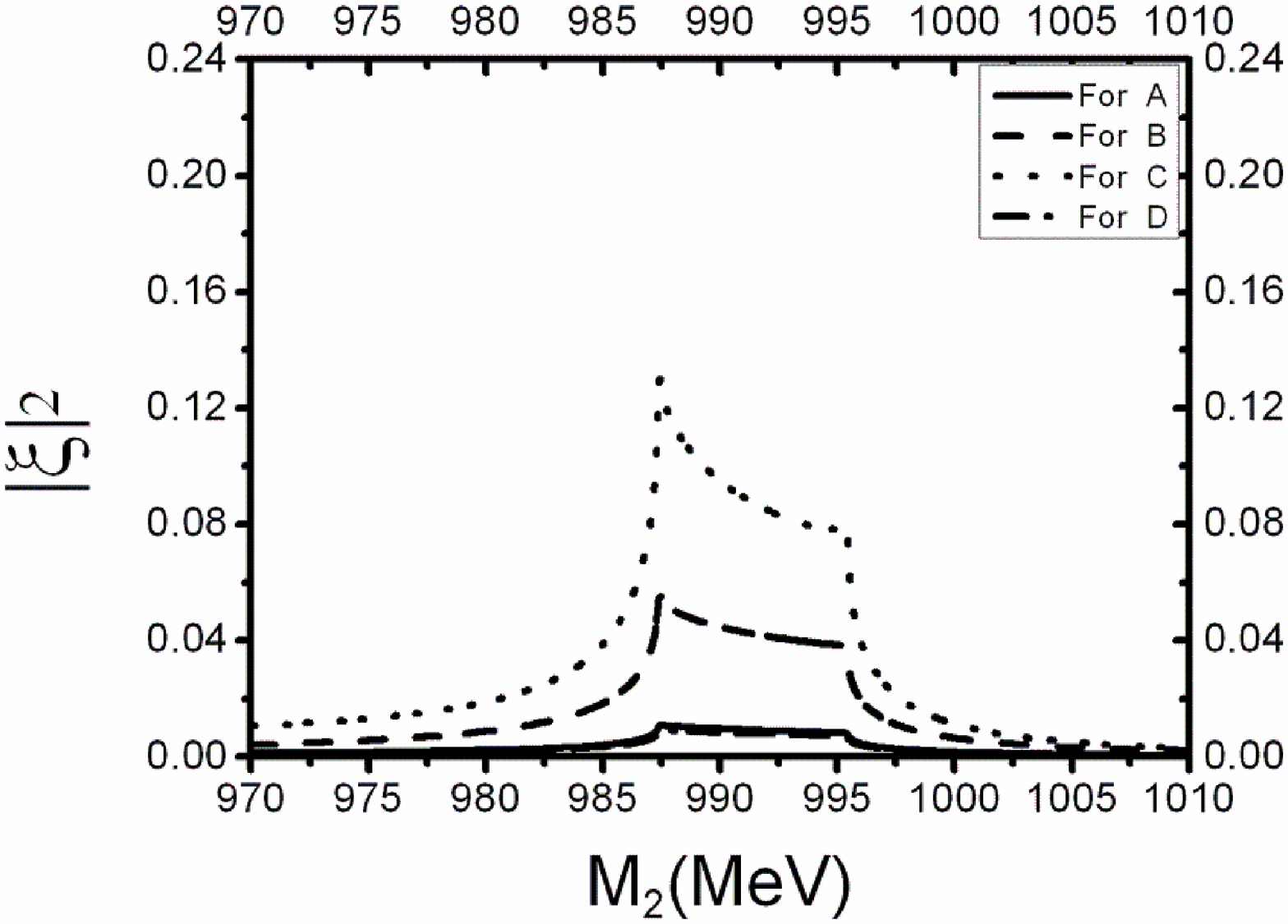}
\includegraphics[width=0.37\columnwidth,angle=270]{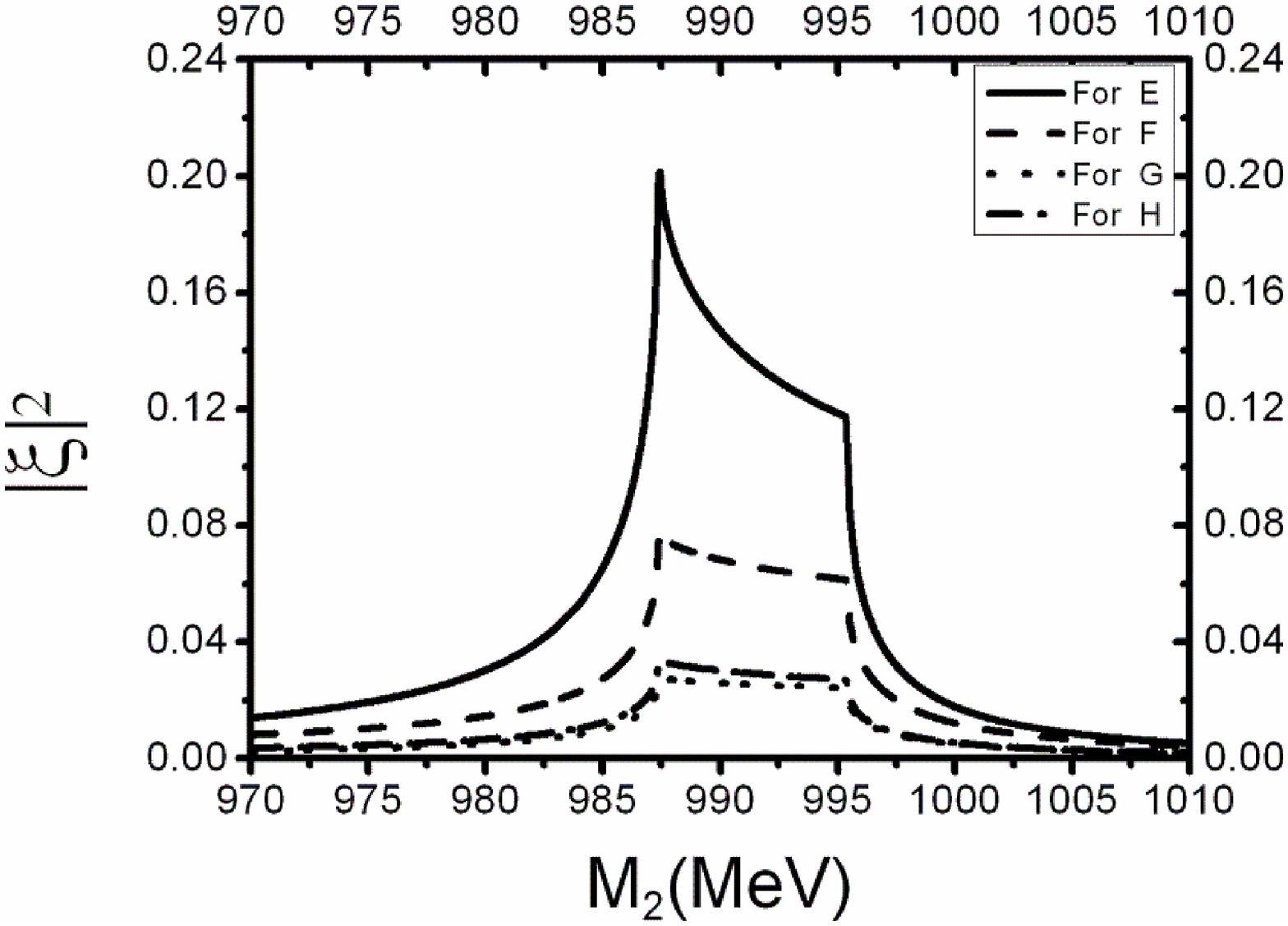}
\caption{Predictions for the \gaf intensity $|\xi|^{2}$ vs
two-meson invariant mass $M_2$ from various models A-D (left) and
various experimental measured coupling constants E-H (right).}
\label{f1}
\end{center}
\end{figure}

From Eqs.(\ref{1}-\ref{5}), one can see that the mixing intensity
$|\xi|^2$ depends on $g_{\ga K^{+}K^{-}}$, $g_{\gf K^{+}K^{-}}$
and $g_{\ga\pi^{0}\eta}$.  Various models for the structures of
$\ga$ and $\gf$ give different predictions for these coupling
constants \cite{eb,m2,m3,m6} as listed in Table \ref{tab-1} by
No.A-D. There have also been some experimental measurements on
these coupling constants \cite{ge1,ge2,ge4,ge5,m5,zou1,zou2} as
listed by No.E-H. The corresponding predictions for the \gaf
intensity $|\xi|^2$ from these various theoretical and
experimental values of the coupling constants are calculated and
plotted  in Fig.\ref{f1}. In the calculation, the masses for
$K^+$, $K^0$, $\pi^0$ and $\eta$ are taken from PDG2006
\cite{pdg06} as $m_{K^{+}}=493.7MeV$, $m_{K^{0}}=497.7MeV$,
$m_{\pi}=135.0 MeV$and $m_{\eta}=547.5MeV$, respectively.

\begin{table}[ht]
\begin{tabular}{|c|c|c|c|c|c|}
\hline No. & model or experiment & $a_{0}$ mass~(MeV) &
$g_{a_{0}\pi\eta}$~(GeV) &
$g_{a_{0}K^{+}K^{-}}$~(GeV) & $g_{f_{0}K^{+}K^{-}}$~(GeV)\\
\hline
 A & $q\bar{q}$ model \cite{eb} & 983 & 2.03 & 1.27 & 1.80\\
\hline
 B & $q^{2}\bar{q}^{2}$ model \cite{eb} & 983 & 4.57 & 5.37 & 5.37\\
\hline
 C & $K\bar{K}$ model \cite{m2,m6} & 980 & 1.74 & 2.74 & 2.74\\
\hline
 D & $q\bar{q}g$ model \cite{m3} & 980 & 2.52 & 1.97 & 1.70\\
\hline
 E & SND \cite{ge1,ge2} & 995 & 3.11 & 4.20 & 5.57\\
\hline
 F & KLOE \cite{ge4,ge5} & 984.8 & 3.02 & 2.24 & 5.92\\
\hline
 G & BNL \cite{m5} & 1001 & 2.47 & 1.67 & 3.26 \cite{zou2}\\
\hline
 H & CB \cite{zou1} & 999 & 3.33 & 2.54 & 4.18 \cite{bes1}\\
\hline
\end{tabular}
\caption{$\ga$ mass and coupling constants $g_{a_{0}\pi\eta}$,
$g_{a_{0}K^{+}K^{-}}$, $g_{f_{0}K^{+}K^{-}}$ from several models
(A-D) and experimental measurements (E-H)} \label{tab-1}
\end{table}

The predictions for $|\xi|^2$ vs $M_2$ peak in the region between
the two thresholds for the charged and neutral kaon systems. The
peak value is in the range of 0.01 to 0.2. It is mainly determined
by the ratio $g_{\ga K^{+}K^{-}}g_{\gf
K^{+}K^{-}}/g^2_{\ga\pi^{0}\eta}$. The different predictions by
various models (No.A-D) indicate that the $f_0-a_0$ mixing depends
on the nature of the scalars with the $K\bar K$ molecule giving
the largest mixing and the four quark state the second. However,
one should keep in mind that the absolute value for the mixing
from each model is quite model-dependent and suffers rather big
uncertainty, which may make it difficult to discriminate between
various pictures as in the case for the radiative decays
$\phi\to\gamma a_0/f_0$ \cite{Moscow}. Nevertheless, a reliable
measurement of the mixing will be very useful to constrain model
parameters and ultimately understand the nature of these scalars.
Present available experimental measurements on the coupling
constants of $g_{\ga K^{+}K^{-}}$, $g_{\gf K^{+}K^{-}}$ and
$g_{\ga\pi^{0}\eta}$ cannot give reliable determination of the
\gaf and hence cannot give much constraint on theoretical models.
Direct precise measurement of the $|\xi|^2$ is needed to provide a
useful check on these model predictions and previous measurements.

\section{Possibility of measuring $a^0_0(980)$-$f_0(980)$ mixing from $J/\psi\to\phi\eta\pi^0$}
\label{s3}

Close and Kirk \cite{im} suggested to study $J/\psi$ decays to the
'forbidden' final states $\omega\ga$ and $\phi\ga$ where they
predicted branching ratios of $O(10^{-5})$. The corresponding
$J/\psi$ to $\phi\gf$ and $\omega\gf$ processes have already been
studied by BESII experiments \cite{bes1,bes2}. Although the two
channels are found to have similar branching ratios, the $\gf$
peak is very outstanding in the $\pi\pi$ invariant mass spectrum
for the $J/\psi\to\phi\pi^+\pi^-$ process \cite{bes1} while it is
much buried by other components in the $J/\psi\to\omega\pi^+\pi^-$
process \cite{bes2}. Therefore the
$J/\psi\to\phi\gf\to\phi\ga\to\phi\eta\pi^0$ is expected to be the
best place for studying \gaf from $J/\psi$ decays.  Due to limited
statistics and relatively poor detection of multi-photon final
states, there is no information available on this channel from
BESII experiment. With the increase of statistics by two orders of
magnitude and much improved photon detection expected at BESIII,
here we give a detailed study on the possibility of measuring the
\gaf from $J/\psi\to\phi\eta\pi^0$ process. Besides the
contribution from the \gaf mechanism, we also examine those from
two background reactions $J/\psi\to\gamma^{*}\to\phi\ga$ and
$J/\psi\to K^{*}\bar{K} + c.c.\to\phi\ga$.

\subsection{Contribution from \gaf in $J/\psi\to\phi\eta\pi^0$ decay}

\begin{figure}[htbp] \vspace{-0cm}
\begin{center}
\includegraphics[width=0.35\columnwidth,angle=270]{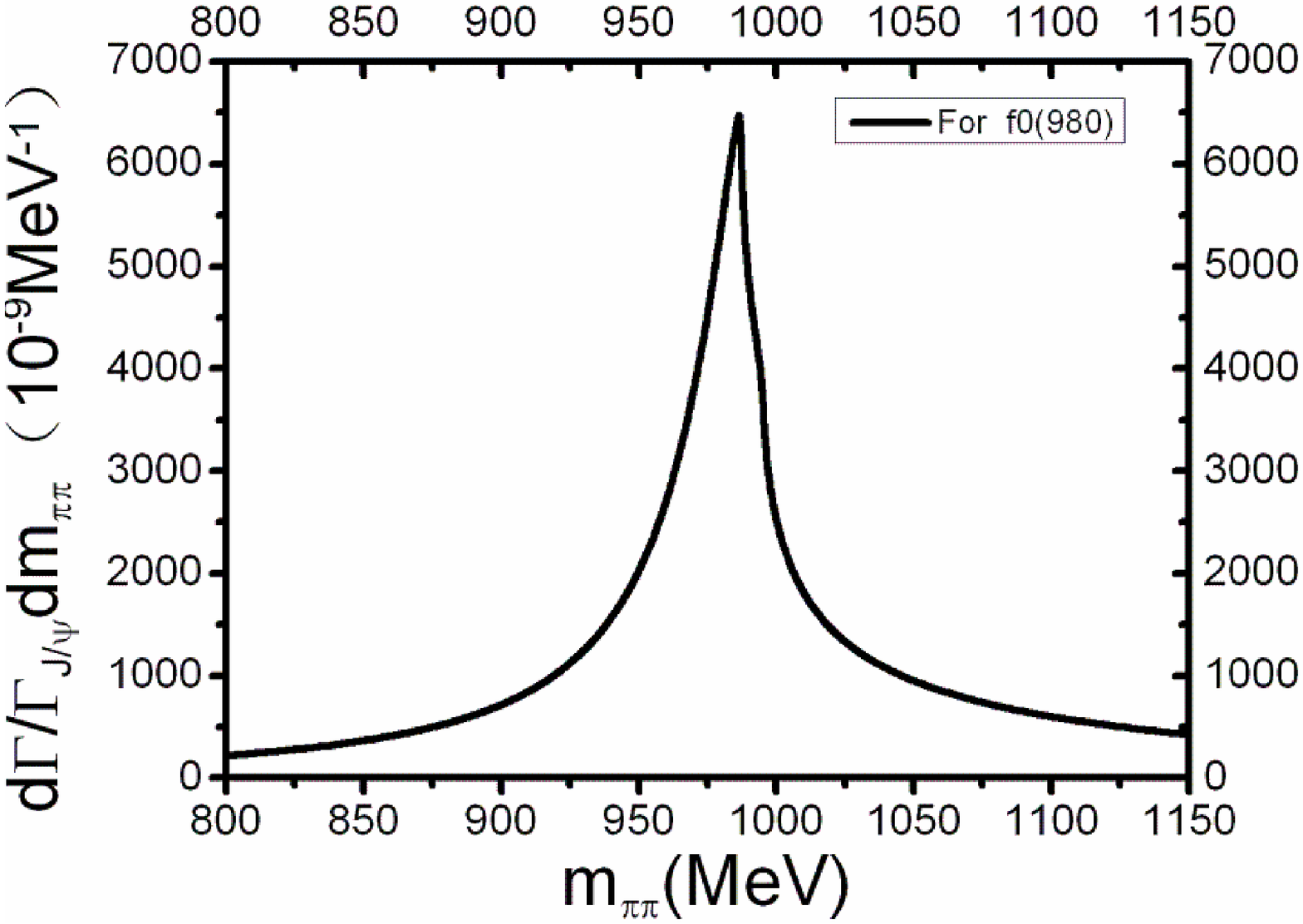}
\includegraphics[width=0.35\columnwidth,angle=270]{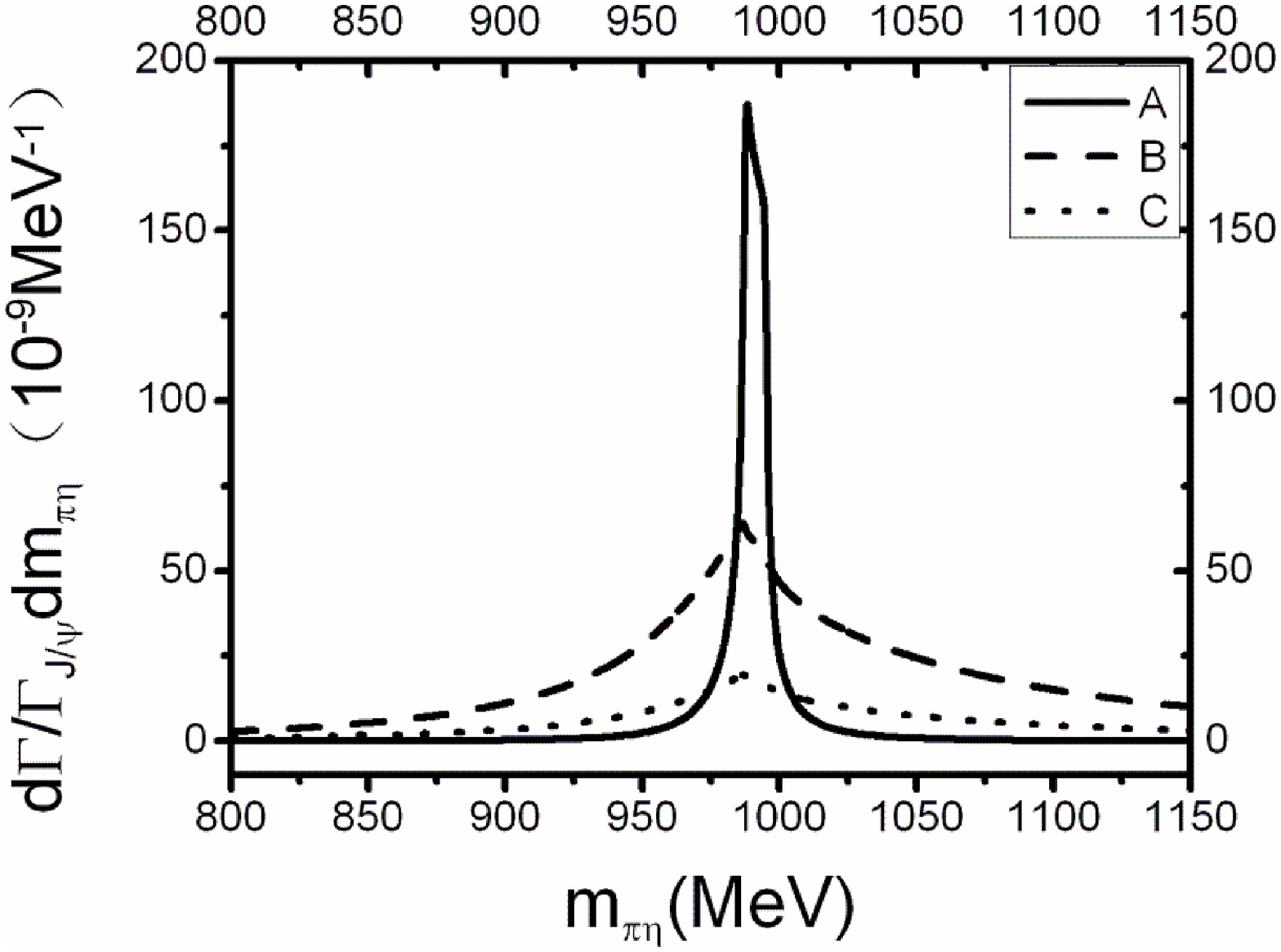}
\caption{$\pi\pi$ invariant mass spectrum for
$J/\psi\to\phi\gf\to\phi\pi\pi$ \cite{bes1} (left) and
corresponding prediction of the $\pi\eta$ invariant mass spectrum
for $J/\psi\to\phi\eta\pi$ through $a_0^0$-$f_0$ mixing (right:
line A) together with estimation of contribution from
$K^{*}\bar{K}$ rescattering (line B for without form factor; C for
monopole form factor with cut-off parameter $\Lambda_K=1.5
GeV$).}\label{f2}
\end{center}
\end{figure}

The observed $\gf$ contribution to the $J/\psi\to\phi\pi\pi$
\cite{bes1} is plotted in Fig.\ref{f2}~(left) with integration
over $m_{\pi\pi}$ equal to the measured branching ratio $(5.4\pm
0.9)\times 10^{-4}$ for this channel. Then from Eq.(\ref{5}) and
parameter set No.H listed in Table 1, we get the corresponding
contribution from \gaf to the $\eta\pi^0$ invariant mass spectrum
for the $J/\psi\to\phi\eta\pi^0$ decay as shown in
Fig.\ref{f2}~(right:~line A). A narrow outstanding peak with a
width about 8 MeV is predicted. A remarkable fact is that the peak
is much narrower than the usual width ($50\sim 100$ MeV) of $\ga$
and should be easily observed even if there are other background
contributions for the $\ga$ production. Integrated over
$m_{\eta\pi}$ for line A, we get the branching ratio about
$2.7\times 10^{-6}$. While parameter set No.G gives a similar
branching ratio, parameter sets No.E and No.F give larger
branching ratio by a factor 5 and 2, respectively.

With $10^9$ $J/\psi$ events and a detection efficiency about 30\%
for the $\phi\eta\pi^0$ channel \cite{lihb} expected at BESIII,
more than 800 events should be observed for this channel with most
events in the narrow gap of $\eta\pi^0$ invariant mass between
987.4 MeV and 995.4 MeV. In the following two subsections we will
show that two background contributions for this channel from
$J/\psi\to\gamma^{*}\to\phi a^{0}_{0(980)}$ and $J/\psi\to
K^{*}\bar{K}+c.c.\to\phi a^{0}_{0(980)}$ as shown in Fig.\ref{f3}
cannot influence the observation of this narrow peak from the \gaf
significantly. Therefore the \gaf intensity is expected to be
unambiguously and precisely measured at BESIII.

\subsection{Contribution from $J/\psi\to\gamma^{*}\to\phi\ga$ to the $\phi\eta\pi^0$ final state}

\begin{figure}[htbp] \vspace{-0.cm}
\begin{center}
\includegraphics[width=0.3\columnwidth,angle=270]{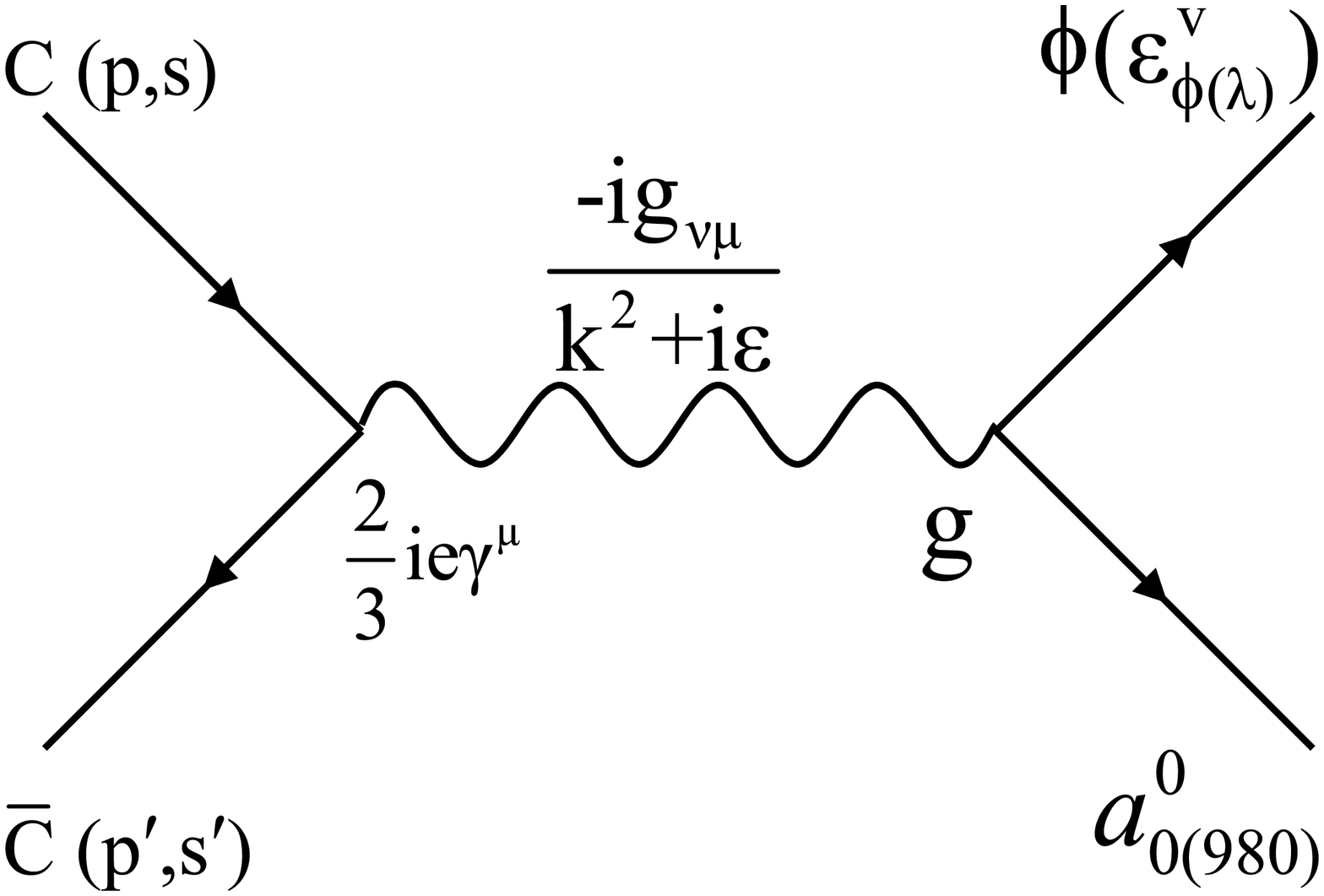}
\hspace{1cm}\includegraphics[width=0.3\columnwidth,angle=270]{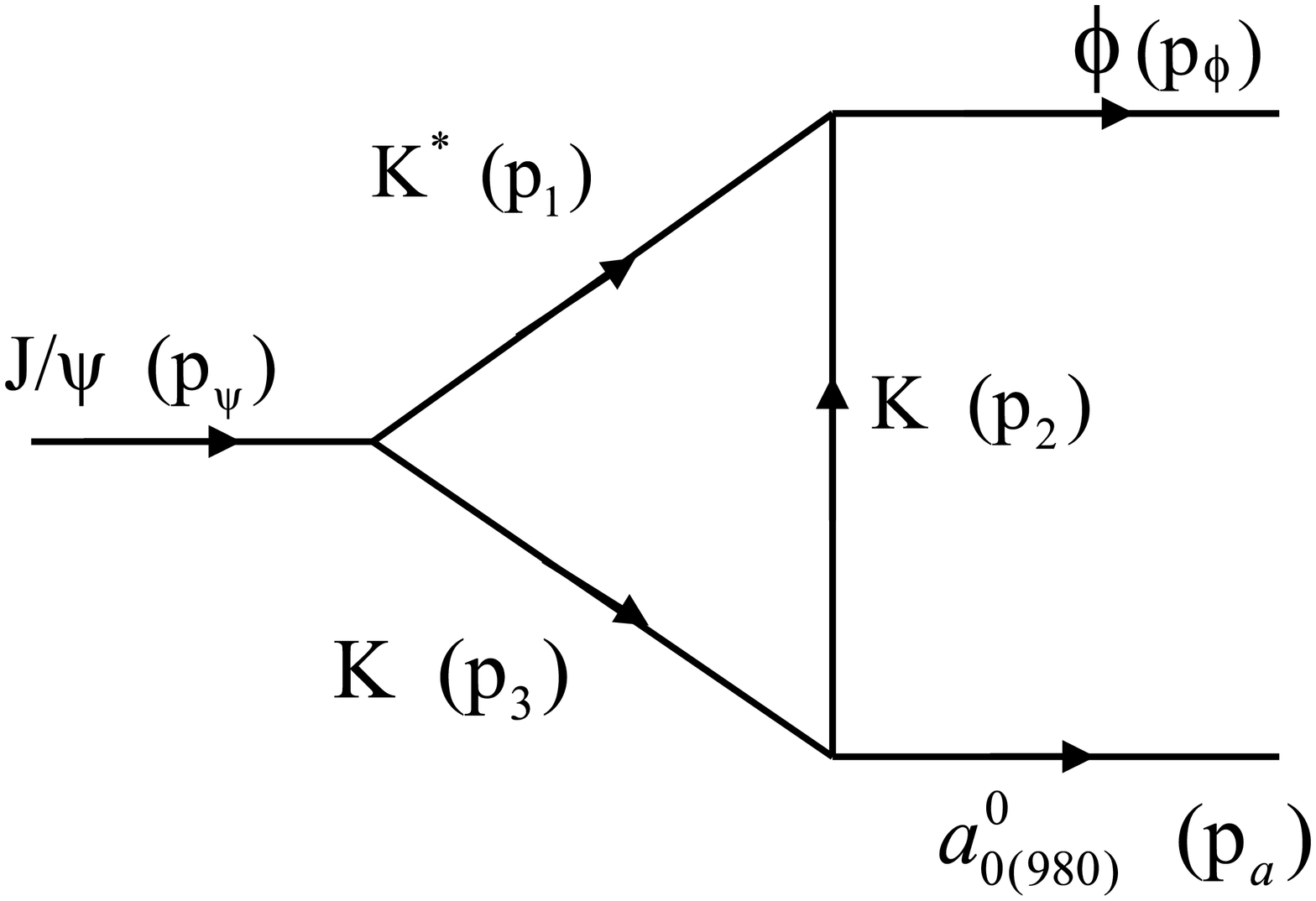}
\caption{Feynman diagrams for reactions
$J/\psi\to\gamma^*\to\phi\ga$ (left) and $J/\psi\to
K^*\bar{K}\to\phi\ga$ (right).}\label{f3}
\end{center}
\end{figure}

Since a common source for the isospin breaking process in $J/\psi$
decays is the electromagnetic decay via $c\bar c$ annihilation to
an intermediate virtual photon, here we examine the contribution
from this mechanism as shown by Fig.\ref{f3}~(left).

The invariant amplitude for $c\bar c\to\gamma^*\to\phi\ga$ is :
\begin{equation}
{\cal M}_{EM} =
\frac{2}{3}ie\bar{u}_{c(p,s)}\gamma^{\mu}v_{\bar{c}(p',s')}
\frac{-ig_{\mu\nu}}{k^{2}+i\varepsilon}g_{\gamma\phi
a(k^2)}e^{\nu}_{\phi(\lambda)},
\end{equation}
where $k$ is the four momentum of $\gamma^{*}$ and $g_{\gamma\phi
a}(k^2)$ is the coupling constant of the virtual photon to
$\phi\ga$. $u_{c(p,s)}$, $v_{\bar{c}(p',s')}$ and
$e^{\mu}_{\phi(\lambda)}$ are spinor wave functions for $c$,
$\bar{c}$ and $\phi$, respectively. Comparing this amplitude with
the invariant amplitude for $c\bar c\to\gamma^*\to e^+ e^-$, the
only difference is a replacement of $ie\bar u_{e^-}\gamma^\nu
v_{e^+}$ by $g_{\gamma\phi a(k^2)}e^{\nu}_{\phi(\lambda)}$.

Then we can easily get the relative ratio of
$J/\psi\to\gamma^{*}\to\phi\ga$ compared to $J/\psi\to\gamma^*\to
e^+ e^-$ as
\begin{equation}
\frac{\Gamma_{J/\psi\to\gamma^{*}\to\phi\ga}}
{\Gamma_{J/\psi\to\gamma^{*}\to e^+e^-}} = \frac{g^2_{\gamma\phi
a(M^2_{J/\psi})}|\vec p_\phi|}{8\pi\alpha
M^3_{J/\psi}}\left[3+\frac{|\vec p_\phi|^2} {m^{2}_{\phi}}\right].
\end{equation}
where $\alpha$ is the electromagnetic fine structure constant,
$M_{J/\psi}$ is the mass of $J/\psi$, $m_\phi$ and $\vec p_\phi$
are mass and momentum of $\phi$ in $J/\psi$ at-rest system,
respectively.

To determine the $\gamma\phi\ga$ coupling constant $g_{\gamma\phi
a(k^2)}$ at $k^2=M^2_{J/\psi}$, we assume a usual monopole form
factor to relate it to its relevant value at $k^2=0$ for a real
photon , {\sl i.e.},
\begin{equation}
g_{\gamma\phi a(k^2)} = g_{\gamma\phi
a(0)}\frac{\Lambda^2}{\Lambda^2-k^2}.
\end{equation}
In the limit of vector meson $\rho^0$ dominance, the parameter
$\Lambda=m_\rho=0.77 GeV$. Considering contributions from other
vector mesons, the $\Lambda$ could be larger to be around 1 GeV.

The $g_{\gamma\phi a(0)}$ can be determined by experimental
information on $\phi\to\gamma a^{0}_{0}$ by the following relation
\begin{eqnarray}
{\cal M}_{\phi\to\gamma a_0} &=& g_{\gamma\phi a(0)}
(e^*_{\phi}\cdot e_{\gamma}), \\
\Gamma_{\phi\to\gamma a_{0}} &=& \frac{g^2_{\gamma\phi
a(0)}}{24\pi m_{\phi}}\left(1-\frac{m^{2}_{a_{0}}}
{m^{2}_{\phi}}\right) .
\end{eqnarray}
From PDG \cite{pdg06} : $Br_{\phi\to\gamma\ga} = 7.6\times
10^{-5}$, $\Gamma_{\phi}=4.26MeV$, $m_{\phi}=1020MeV$,
$m_{a_{0}}=985MeV$, we get $g_{\gamma\phi a(0)} \approx 19.2 MeV
$. Then from Eq.(12) with $\Lambda=1 GeV$ and $M_{J/\psi}=3.1 GeV$
we have $g_{\gamma\phi a(M^2_{J/\psi})}\approx -2.23 MeV$. This is
substituted into Eq.(11) and results in
\begin{equation}
\frac{\Gamma_{J/\psi\to\gamma^{*}\to\phi\ga}}
{\Gamma_{J/\psi\to\gamma^{*}\to e^+e^-}} \approx 4.67\times
10^{-6}.
\end{equation}
With known branching ratio $BR(J/\psi\to e^+e^-)=5.55\%$
\cite{pdg06}, we obtain the branching ratio for
$J/\psi\to\gamma^{*}\to\phi\ga$ as $2.59\times 10^{-7}$. This is
much smaller than the contribution from \gaf and is distributed in
much large range of $\eta\pi^0$ invariant mass spectrum. So its
influence to the narrow \gaf peak is negligibly small.

\subsection{Contribution from $J/\psi\to K^{*}\bar{K}+c.c.\to\phi\ga$
to the $\phi\eta\pi^0$ final state}

Another possible source for the isospin breaking process is due to
$K^*\bar K + c.c.$ rescattering as shown in Fig.\ref{f3}~(right).
The transition amplitude can be expressed as
$(K=K^{+},K^{-},K^{0},\bar{K}^{0})$ :
\begin{equation}
{\cal M}_{FSI}=i\sum_{K}\int\frac{d^4 p_{2(K)}}{(2\pi)^4}T_{\phi
(K)}^{\beta}
\left(-g_{\beta\lambda}+\frac{p_{1(K)\beta}p_{1(K)\lambda}}{p_{1(K)}^{2}}\right)
T_{\psi(K)}^\lambda T_{a(K)}\frac{F(p^2_{2(K)})}{a_{1(K)} a_{2(K)}
a_{3(K)}},
\end{equation}
where denominators of meson propagators are
$a_1=p_1^2-m_1^2+i\epsilon$, $a_2=p_2^2-m_2^2 +i\epsilon$, $a_3 =
p_3^2-m_3^2 +i\epsilon$;
$T_\phi^\beta=i\left(g_\phi\epsilon^{\mu\nu\alpha\beta}
p_{\phi\mu}e_{\phi\nu}p_{2\alpha}\right)/m_\phi$,
$T^\lambda_\psi=i\left(g_\psi \epsilon^{\lambda\sigma\tau\delta}
p_{\psi\sigma}e^*_{\psi\tau} p_{3\delta}\right)/M_{J/\psi}$,
$T_{a}=ig_{a}$ are effective interactions at each vertex.
$F(p^2_2)$ is the off-shell form factor. Three effective coupling
constants $g_\psi$, $g_\phi$ and $g_{a}$ can be determined
independently in relevant meson decays as follows.

The coupling constant $g_{\psi(K)}$ can be determined by the
corresponding decay width of $J/\psi \to K^*\bar{K}+c.c.$
\cite{zhao}:
\begin{eqnarray}
g_{\psi(K^+)}^2=g_{\psi(K^-)}^2=\frac{12\pi M_{J/\psi}^2}{| \vec
p_{1(K^+)}|^3}\Gamma^{exp}_{J/\psi\to K^{*-}K^{+}},\ \
g_{\psi(K^0)}^2=g_{\psi(\bar K^0)}^2=\frac{12\pi
M_{J/\psi}^2}{|\vec p_{1(K^0)}|^3}\Gamma^{exp}_{J/\psi\to
\bar{K}^{*0}K^{0}},
\end{eqnarray}
where $\Gamma^{exp}_{J/\psi\to
K^{*-}K^{+}}/\Gamma_{J/\psi}=(2.5\pm 0.2)\times10^{-3}$,
$\Gamma^{exp}_{J/\psi\to
\bar{K}^{*0}K^{0}}/\Gamma_{J/\psi}=(2.1\pm 0.2)\times10^{-3}$ and
$\Gamma_{J/\psi}=93.4\pm 2.1 keV$ from PDG \cite{pdg06}.

For $g_{\phi}$, because $\omega$ and $\phi$ are nearly ideally
mixed, the SU(3) symmetry leads to $g_\phi=g_{\omega
\rho^0\pi^0}/\sqrt{2}$, where $g^2_{\omega \rho^0\pi^0}\simeq 84$
determined by $\omega\to\rho\pi\to 3\pi$ decay width \cite{zhao}.

For the $g_a$ coupling, we have $g_a\equiv
g_{a(K^+)}=g_{a(K^-)}=-g_{a(K^0)}=-g_{a(\bar K^0}=2.54 GeV$ from
Ref.\cite{zou1,zou2}.

For the loop calculation in Eq.(16), we assume the on-shell
approximation by applying the Cutkosky rule as in
Refs.\cite{zhao,locher}, then the transition amplitude reduces to
\begin{equation}
{\cal M}_{FSI}=-i\sum_{K}\frac{|\vec{p}_{3(K)}|^{2}
g_{\phi}g_{\psi(K)}g_{a}} {32\pi^{2}M_{J/\psi}^{2}m_\phi}\int
d\Omega_{p_3} \frac{T_{(K)}F(p^2_{2(K)})} {p^2_{2(K)}-m^2_{2(K)}}
.
\end{equation}
And
\begin{eqnarray}
|{\cal M}_{FSI}|^{2}=
A\left|U_{(K^+)}-U_{(K^0)}\right|^{2}=A\left|U_{(K^+)}^{2}+U_{(K^0)}^{2}-2U_{(K^+)}U_{(K^0)}\right|,
\end{eqnarray}
where
\begin{eqnarray}
A&=&\frac{4\left(g_\phi g_a\right)^2}{3\left(32\pi^2M_{J/\psi}m_\phi\right)^{2}},\\
T_{(K)}&=&\varepsilon^{\mu\nu\alpha\beta}\varepsilon_{\beta\sigma\tau\delta}p_{\phi\mu}
e_{\phi\nu}p_{2(K)\alpha}
p_\psi^{\sigma}e_\psi^{*\tau}p_{3(K)}^{\delta},\\
U_{(K)}&=&\int d\Omega_{p_3} \frac{g_{\psi(K)}|{\vec
p}_{3(K)}|T_{(K)} F(p^2_{2(K)})}{p^2_{2(K)}-m^2_{2(K)}},\ \ \
(K=K^0,K^+)
\end{eqnarray}
with $p_2=p_a-p_3$. The form factor $F(p^2_2)$ is included to take
into account the off-shell effects for the exchanged meson in the
final state interactions.

Define
\begin{eqnarray}
B_{(K)} &= & 2|\vec p_\phi||\vec p_{3(K)}| /(M_{a}^2+m_{3(K)}^2-2E_{a}E_{3(K)}-m_{2(K)}^{2}),\\
C_{(K)}&= & 2|\vec p_\phi||\vec p_{3(K)}|
/(M_{a}^2+m_{3(K)}^2-2E_{a}E_{3(K)}-\Lambda_K^2),
\end{eqnarray}
then we get $U_{(K)}U_{(K')}$ for various form factor $F(p^2_2)$
as the following.

\noindent (i) Without form factor: $F(p^2_2)=1$
\begin{eqnarray}
U_{(K^{+})}^{2}&=&\frac{\pi^{2}g^{2}_{\psi(K^{+})}B_{(K^{+})}^{2}}
{|\vec{p}_{\phi}|^{2}}\int^{1}_{-1}dtdt'\frac{T_{(K^{+})}T'_{(K^{+})}}
{(1+B_{(K^{+})}t)(1+B_{(K^{+})}t')},\\
U_{(K^{0})}^{2}&=&\frac{\pi^{2}g^{2}_{\psi(K^{0})}B_{(K^{0})}^{2}}
{|\vec{p}_{\phi}|^{2}}\int^{1}_{-1}dtdt'\frac{T_{(K^{0})}T'_{(K^{0})}}
{(1+B_{(K^{0})}t)(1+B_{(K^{0})}t')},\\
U_{(K^{+})}U_{(K^{0})}&=&\frac{\pi^{2}g_{\psi(K^{+})}g_{\psi(K^{^{0}})}
B_{(K^{+})}B_{(K^{0})}}{|\vec{p}_{\phi}|^{2}}\int^{1}_{-1}dtdt'\frac{T_{(K^{0})}T'_{(K^{+})}}
{(1+B_{(K^{0})}t)(1+B_{(K^{+})}t')},
\end{eqnarray}
(ii) Monopole form factor:
$F(p^2_2)=(\Lambda_K^{2}-m_{2(K)}^{2})/(\Lambda_K^{2}-p_{2(K)}^{2})$
\begin{eqnarray}
U_{(K^{+})}^{2}&=&\frac{\pi^{2}g^{2}_{\psi(K^{+})}B_{(K^{+})}^{2}}
{|\vec{p}_{\phi}|^{2}}\frac{C_{(K^{+})}^{2}(m^{2}_{2(K^{+})}-\Lambda^2_K)^{2}}
{4|\vec{p}_{\phi}|^{2}|\vec{p}_{1(K^{+})}|^{2}}\nonumber\\
&&\times\int^{1}_{-1}dtdt'\frac{T_{(K^{+})}T'_{(K^{+})}}{(1+B_{(K^{+})}t)
(1+B_{(K^{+})}t')(1+C_{(K^{+})}t)(1+C_{(K^{+})}t')},\\
U_{(K^{0})}^{2}&=&\frac{\pi^{2}g^{2}_{\psi(K^{0})}B_{(K^{0})}^{2}}
{|\vec{p}_{\phi}|^{2}}\frac{C_{(K^{0})}^{2}(m^{2}_{2(K^{0})}-\Lambda^2_K)^{2}}
{4|\vec{p}_{\phi}|^{2}|\vec{p}_{1(K^{0})}|^{2}}\nonumber\\
&&\times\int^{1}_{-1}dtdt'\frac{T_{(K^{0})}T'_{(K^{0})}}{(1+B_{(K^{0})}t)
(1+B_{(K^{0})}t')(1+C_{(K^{0})}t)(1+C_{(K^{0})}t')},\\
U_{(K^{+})}U_{(K^{0})}&=&\frac{\pi^{2}g_{\psi(K^{+})}g_{\psi(K^{0})}B_{(K^{+})}B_{(K^{0})}}
{|\vec{p}_{\phi}|^{2}}\frac{C_{(K^{+})}C_{(K^{0})}(m^{2}_{2(K^{0})}-\Lambda^2_K)(m^{2}_{2(K^{+})}
-\Lambda^2_K)}{4|\vec{p}_{\phi}|^{2}|\vec{p}_{1(K^{+})}||\vec{p}_{1(K^{0})}|}\nonumber\\
&&\times\int^{1}_{-1}dtdt'\frac{T_{(K^{0})}T'_{(K^{+})}}{(1+B_{(K^{0})}t)(1+B_{(K^{+})}t')
(1+C_{(K^{0})}t)(1+C_{(K^{+})}t')},
\end{eqnarray}
iii) Dipole form factor:
$F(p^2_2)=[(\Lambda_K^{2}-m_{2(K)}^{2})/(\Lambda_K^{2}-p_{2(K)}^{2})]^2$
\begin{eqnarray}
U_{(K^{+})}^{2}&=&\frac{\pi^{2}g^{2}_{\psi(K^{+})}B_{(K^{+})}^{2}}{|\vec{p}_{\phi}|^{2}}
\left(\frac{C_{(K^{+})}^{2}(m^{2}_{2(K^{+})}-\Lambda^2_K)^{2}}
{4|\vec{p}_{\phi}|^{2}|\vec{p}_{1(K^{+})}|^{2}}\right)^{2}\nonumber\\
&&\times\int^{1}_{-1}dtdt'\frac{T_{(K^{+})}T'_{(K^{+})}}{(1+B_{(K^{+})}t)(1+B_{(K^{+})}t')
(1+C_{(K^{+})}t)^{2}(1+C_{(K^{+})}t')^{2}},\\
U_{(K^{0})}^{2}&=&\frac{\pi^{2}g^{2}_{\psi(K^{0})}B_{(K^{0})}^{2}}{|\vec{p}_{\phi}|^{2}}
\left(\frac{C_{K^{0}}^{2}(m^{2}_{2(K^{0})}-\Lambda^2_K)^{2}}
{4|\vec{p}_{\phi}|^{2}|\vec{p}_{1(K^{0})}|^{2}}\right)^{2}\nonumber\\
&&\times\int^{1}_{-1}dtdt'\frac{T_{(K^{0})}T'_{(K^{0})}}{(1+B_{(K^{0})}t)(1+B_{(K^{0})}t')
(1+C_{(K^{0})}t)^{2}(1+C_{(K^{0})}t')^{2}},\\
U_{(K^{+})}U_{(K^{0})}&=&\frac{\pi^{2}g_{\psi(K^{^{+}})}g_{\psi(K^{0})}B_{(K^{+})}B_{(K^{0})}}
{|\vec{p}_{\phi}|^{2}}\left(\frac{C_{(K^{+})}C_{(K^{0})}(m^{2}_{2(K^{+})}-\Lambda^2_K)(m^{2}_{2(K^{0})}
-\Lambda^2_K)}{4|\vec{p}_{\phi}|^{2}|\vec{p}_{1(K^{0})}||\vec{p}_{1(K^{+})}|}\right)^{2}\nonumber\\
&&\times\int^{1}_{-1}dtdt'\frac{T_{(K^{0})}T'_{(K^{+})}}{(1+B_{(K^{0})}t)(1+B_{(K^{+})}t')
(1+C_{(K^{0})}t)^{2}(1+C_{(K^{+})}t')^{2}}.
\end{eqnarray}

For the final calculation of $J/\psi\to
K^{*}\bar{K}+c.c.\to\phi\ga\to\phi\eta\pi^0$, we include the
$g_{a_0\eta\pi}/D_a$ factor with parameter set No.H for the $\ga$
propagator.  Taking into account also $\ga\to K\bar K$ we get the
branching ratio of $J/\psi\to\phi\ga$ for monopole and dipole form
factors with typical $\Lambda_K$ cut-off parameters as listed in
the Table \ref{tab-2}. For $\Lambda_K=\infty$, it is equivalent to
without form factor, {\sl i.e.}, $F(p^2_2)=1$.

\begin{table}[ht]
\begin{tabular}{|c|c|c|}
\hline
 $\Lambda_K$ (GeV)  & monopole F.F. & dipole F.F. \\
\hline
 1.0 &  $1.5\times 10^{-6}$ &  $0.4\times 10^{-6}$ \\
\hline
 1.5 &  $3.8\times 10^{-6}$ &  $2.1\times 10^{-6}$ \\
\hline
 2.0 &  $5.7\times 10^{-6}$ &  $4.6\times 10^{-6}$ \\
\hline
 $\infty$ & $12.3\times 10^{-6}$ & $12.3\times 10^{-6}$ \\
\hline
\end{tabular}
\caption{Branching ratio of $J/\psi\to K^*\bar{K}\to\phi\ga$ for
monopole and dipole form factors with typical $\Lambda_K$ cut-off
parameters. }\label{tab-2}
\end{table}

For the most commonly used monopole form factor with
$\Lambda_K=1.5 GeV$, the branching ratio from on-shell $K^*\bar
K+c.c.$ rescattering is about $3.8\times 10^{-6}$. The
corresponding off-shell loop is expected to give a similar amount
of contribution. So the branching ratio from $K^*K$ loops could be
a few times more than that from the \gaf for the
$J/\psi\to\phi\ga$. However as shown by line B and line C in
Fig.\ref{f2}, the contribution from $K^*K$ loops gives a much
broader distribution in the $\pi\eta$ invariant mass spectrum than
that from the \gaf. Although the integration of line B is about
3.5 times of line A, the peak of line A is still more than a
factor of 2 over the peak of line B. Therefore by separate the
narrow peak from the broader peak, we can still get very precise
measurement for the \gaf.

\section{Summary}
\label{s4}

The predictions of \gaf intensity $|\xi|^2$ from various models,
such as $q\bar q$, $q^2\bar q^2$, $K\bar K$ and $q\bar q g$ for
$\ga$ and $\gf$, are summarized and shown in Fig.\ref{f1} (left)
with a range of $0.01\sim 0.1$.  The deduced \gaf intensity from
various measurements of the relevant coupling constants is
summarized and shown in Fig.\ref{f1} (right) with a range of
$0.02\sim 0.2$. This large uncertainty is not good enough to
distinguish various models. More solid and precise measurement of
this quantity is needed.

In this paper, we examine the possibility of measuring the \gaf
from $J/\psi\to\phi\gf\to\phi\ga\to\phi\eta\ga$ reaction at
upgraded Beijing Electron Positron Collider with BESIII detector.
We find that the \gaf gives a branching ratio of $O(10^{-6})$ to
the $J/\psi\to\phi\ga$ and a narrow peak about 8 MeV at about 990
MeV in the $\eta\pi^0$ invariant mass spectrum. The contribution
from $J/\psi\to\gamma^{*}\to\phi\ga$ is negligibly small. The
contribution from $J/\psi\to K^{*}\bar{K}+c.c.\to\phi\ga$ also
gives a branching ratio of $O(10^{-6})$, but with a much broader
width about $50\sim 100$ MeV which should be easily separated from
the narrow structure caused by the \gaf. With $10^9$ $J/\psi$
events and a detection efficiency about 30\% for the
$\phi\eta\pi^0$ channel \cite{lihb} expected at BESIII, for
$|\xi|^2$ in the range of $0.01\sim 0.2$, in should be easily
measured with a precision $\Delta|\xi|^2/|\xi|^2<10\%$.

Finally we want to give a comment on the result
$|\xi|^{2}=(8\pm3)\%$ by Ref.\cite{im}. The result is based on the
data \cite{WA102} of the $\ga$ central production in the reaction
$pp\to p_s(\eta\pi^0)p_f$ and assumes that the $\ga$ peak comes
from the \gaf. However, the width of $\ga$ peak in the $\eta\pi^0$
invariant mass spectrum is found to be $72\pm 16$ MeV similar to
the width of $a^-_0(980)$ peak in the $\eta\pi^-$ invariant mass
spectrum as $61\pm 19$ MeV in the WA102 experiment \cite{WA102}.
Since the \gaf is shown in Fig.\ref{f2} to give a much narrower
width of about 8 MeV, the $\ga$ peak from WA102 experiment is
unlikely mainly coming from the \gaf mechanism.

\bigskip
\noindent {\bf Acknowledgements} We would like to thank H.B.Li,
G.Li and Ulf-G. Meissner for useful discussions. This work is
partly supported by the National Natural Science Foundation of
China (NSFC) under grants Nos. 10435080, 10521003, 10675131 and by
the Chinese Academy of Sciences under project No. KJCX3-SYW-N2.

\end{document}